\documentclass[11pt,epsf]{article}
\usepackage{amsmath}
\usepackage{amsfonts}
\usepackage{amssymb}
\usepackage{graphicx}
\usepackage{color}
\usepackage{comment}

\newtheorem{theorem}{Theorem}

\newcommand {\dfn} {\stackrel{\Delta} {=}}
\newcommand {\exe} {\stackrel{\cdot} {=}}
\newcommand{\eqa}{\stackrel{\mbox{\tiny (a)}}{=}}
\newcommand{\eqb}{\stackrel{\mbox{\tiny (b)}}{=}}
\newcommand{\eqc}{\stackrel{\mbox{\tiny (c)}}{=}}

\newcommand{\eqe}{\stackrel{\mbox{\tiny (e)}}{=}}

\newcommand{\led}{\stackrel{\mbox{\tiny (d)}}{\le}}

\newcommand {\reals} {{\rm I\!R}}

\newcommand {\bs} {\mbox{\boldmath $s$}}

\newcommand {\bw} {\mbox{\boldmath $w$}}

\newcommand {\bE} {\mbox{\boldmath $E$}}

\newcommand{\calH}{{\cal H}}
\newcommand{\calI}{{\cal I}}

\newcommand{\calN}{{\cal N}}

\newcommand{\calS}{{\cal S}}

\allowdisplaybreaks

\begin{document}
\thispagestyle{empty}
\title{Optimal Correlators and Waveforms for Mismatched Detection}
\author{Neri Merhav}
\date{}
\maketitle

\begin{center}
The Andrew \& Erna Viterbi Faculty of Electrical and Computer Engineering\\
Technion - Israel Institute of Technology \\
Technion City, Haifa 32000, ISRAEL \\
E--mail: {\tt merhav@ee.technion.ac.il}\\
\end{center}
\vspace{1.5\baselineskip}
\setlength{\baselineskip}{1.5\baselineskip}

\begin{abstract}
We consider the classical Neymann--Pearson hypothesis testing problem of signal
detection, where under the null hypothesis ($\calH_0$), the received signal is
white Gaussian noise, and under the alternative hypothesis ($\calH_1$), the received
signal includes also an
additional non--Gaussian random
signal, which in turn
can be viewed as a deterministic waveform plus zero--mean, non-Gaussian
noise. However, instead of the classical
likelihood ratio test detector, which might be difficult to implement, in general,
we impose a (mismatched) correlation detector, which is relatively easy to implement, and
we characterize the optimal correlator weights in the sense of the best trade-off
between the false-alarm error exponent and the missed-detection error
exponent. Those optimal correlator weights depend (non-linearly, in general) on the underlying
deterministic waveform under $\calH_1$. We then assume that the deterministic
waveform may also be free to be optimized (subject to a power constraint),
jointly with the correlator, and show that both the optimal waveform and the
optimal correlator weights may take on values in a small finite set of
typically no more than two to four levels, depending on the distribution of
the non-Gaussian noise component. Finally, we outline an extension of the
scope to a wider class of detectors that are based on linear combinations of
the correlation and the energy of the received signal.\\

\end{abstract}

\section{Introduction}

The topic of detection of signals corrupted by noise has a very long history
of active research efforts, as it has an extremely
wide spectrum of engineering applications in the areas communications and signal
processing. These include radar,
sonar, light detection and ranging (LIDAR),
object recognition in images
and video streams,
diagnosis based on biomedical signals,
watermark detection in images and audio signals, 
seismological signal detection related to geophysical activity, and
object detection using multispectral/hyperspectral imaging,
just to name a few. One of the most problematic and frequently encountered issues in signal detection
scenarios is mismatch between the signal model and
the detector design, which is based upon certain assumptions on that model.
Accordingly, the topic of mismatched signal detection has received considerable attention in the literature, see, e.g., 
\cite{BOR09}, \cite{GGFL98}, \cite{HLYC20}, \cite{LL19}, \cite{LLGWW19},
\cite{LXLGHW15}, \cite{LXW14}, \cite{WLYTDY15}, and
\cite{ZGJJ20}, for a non-exhaustive list of relevant references. The common
theme in most of these works is the possible presence of uncertainties in the
desired signal to be detected, in the steering vector, in the transfer function of
the propagation medium, and/or in the distributions of the various kinds of
noise, interference and clutter. Accordingly, adaptive detection mechanisms
with tunable parameters have been developed and proposed in order to combat those types of
mismatch. 

Another line of earlier relevant research activity is associated
with the notion of {\it robust detection techniques}, where the common theme
is generally directed towards a worst-case design of the detector against
small non-parametric uncertainties around
some nominal noise distribution, most notably, a Gaussian distribution. See, e.g., \cite{Capon61}, \cite{E-SV77},
\cite{E-SV79}, \cite{Geraniotis85}, \cite{Kassam81}, \cite{KMS82}, \cite{KT76}, \cite{Kay82},
\cite{Krasnenker80}, \cite{MS71}, \cite{Moustakides85}, and \cite{MT84}.
See also \cite{KP85} for a survey on the subject. 

Last but not least,
when the uncertainty is only in a finite number of parameters of the model, the problem
is normally treated in the framework of composite hypothesis testing, where the popular
approach is the well--known generalized likelihood ratio test (GLRT)
\cite{vanTrees}, which is
often (but not always) asymptotically optimal in the error exponent sense,
see, for example, \cite{CR98}, \cite{EF00a}, \cite{EF00b}, and \cite{ZZM92}.
The GLRT is applied also in
some of the above cited articles on mismatched detection, among many
others. Another approach to composite hypothesis testing is the competitive minimax approach,
proposed in \cite{FM02}.

Our objective in this work is partially related to those studies, but it is different.
It is associated with mismatched detection, except that the origin of this mismatch is
not quite due to uncertainty in the signal-plus-noise model, but it comes from practical
considerations: the optimal likelihood ratio test (LRT) detector might be
difficult to implement in many application examples, especially in the case
of sensors that are built on small, mobile devices which are subjected to severe limitations on
power and computational resources. In such situations, it is desirable that
the detector would be as simple as possible, e.g., a correlation detector, or
a detector that is based on correlation and energy. Within this framework, the
number of arithmetic operations (especially the multiplications) should be made as
small as possible. Clearly, a detector from this class cannot be optimal,
unless the noise is Gaussian, hence the mismatch. Nonetheless, we would like
to find the best correlator weights in the sense of optimizing the trade-off
between the false-alarm (FA) and the missed--detection (MD) rates. This would
partially compensate for the mismatch in case the noise is not purely
Gaussian.

More precisely, consider the following signal detection problem, of distinguishing between two
hypotheses:
\begin{eqnarray}
& &\calH_0:~~Y_t=N_t,~~~~~~t=1,2,\ldots,n\\
& &\calH_1:~~Y_t=X_t+N_t,~~~~~~~t=1,2,\ldots,n
\end{eqnarray}
where $\{N_t\}$ is an independently identically distributed (IID), zero-mean Gaussian noise process with variance $\sigma_N^2$,
independent of $\{X_t\}$, which is another random process, that we decompose
as $X_t=s_t+Z_t$, with $s_t=\bE\{X_t\}$ being a deterministic waveform and
$Z_t=X_t-s_t$ being an IID, zero-mean noise process, which is not necessarily Gaussian in
general. The non--Gaussian noise component, $\{Z_t\}$, can be thought of as
signal--induced noise (SIN), which may stem from several possible mechanisms, such
as: echos of the desired signal, multiplicative noise, cross-talks from parallel channels conveying
correlated signals, interference by jammers, and in the case of optical
detection using avalanche photo-diodes (APDs), it corresponds to shot noise plus
multiplicative noise due to the random gain of the device (see, e.g., \cite{me21} and
references therein, for more details). In general, $\{Z_t\}$ may also
designate randomness that could be attributed to uncertainty associated with the
transmitted signal.

As mentioned above, the optimal LRT detector might
be considerably difficult to implement in practice
since the probability density function
(PDF) of $\{Y_t\}$ under $\calH_1$ involves the convolution between the Gaussian PDF of $N_t$ and
the (non-Gaussian) PDF of $Z_t$, which is typically complicated. As said, a reasonable
practical compromise, valid when the underlying signal $\{s_t\}$ is not
identically zero, is 
a correlation detector, which compares the correlation, $\sum_{t=1}^nw_tY_t$, to 
a threshold, where $w_1,\ldots,w_n$ are referred to as the {\em correlator
weights}, and the threshold
controls the trade-off between the FA probability
and the MD probability. Our first objective is to
characterize the best correlator weights, $w_1^*,\ldots,w_n^*$, in the sense
of the optimal trade-off between the FA probability and the MD probability, or more
precisely, the optimal trade-off between the asymptotic exponential rates of
decay of these probabilities as functions of the sample size $n$, i.e., the
{\em FA exponent} and the {\em MD exponent}. Clearly, the optimal correlation
detector is, in general, not as good as the optimal, LRT detector, but it is
the best compromise between performance and the practical implementability in
the framework of correlation detectors.
As very similar study was
already carried out in \cite{me21}, in the context of optical signal detection
using photo-detectors, where the optimal correlator waveform was characterized in
terms of the optical transmitted signal in continuous time, and was found to be
given by a certain non-linear function of the optical signal.

Here, we study the problem in a more general framework, in the sense that the
PDF of the SIN, $Z_t$, is arbitrary. Moreover, we expand the scope in several directions, in
addition to the study that is directly parallel to that of \cite{me21}.
\begin{enumerate}
\item We consider the possibility of limiting the number of levels $\{w_t\}$ to be finite (e.g., binary,
ternary, etc.), with the motivation of significantly reducing the
number of multiplications needed to calculate the correlation, $\sum_tw_tY_t$.
\item We jointly optimize
both the signal $\{s_t\}$ and the correlator,
$\{w_t\}$. Interestingly, here both the optimal signal and the optimal
correlation weights turn out to have a {\em finite number of levels} even if this
number is not restricted a-priori. The number of levels depends on the
PDF of $Z_t$, and it is typically very small (e.g., two to four levels). Moreover, the
optimal $\{s_t\}$ and $\{w_t\}$ turn out to be proportional to each other, in contrast to
the non-linear relation resulting when only $\{w_t\}$ is optimized while
$\{s_t\}$ is given. 
\item We outline an extension to a wider class of detectors that are based on
linear combinations of the correlation, $\sum_tw_tY_t$, and the energy,
$\sum_tY_t^2$, with the motivation that it is, in fact, the structure of the
optimal detector when $Z_t$ is Gaussian noise, and that it is reasonable
regardless, since the under $\calH_1$ the power (or the variance) of the received signal is
larger than under $\calH_0$.\footnote{In fact, when $s_t\equiv 0$, the
correlation term becomes useless altogether and the energy term becomes necessary.}
We also address the possibility of replacing the energy term by the
sum of absolute values, $\sum_t|Y_t|$, which is another measure of signal
intensity, with the practical advantage that its calculation does not require multiplications.
\end{enumerate}

The outline of the remaining part of this work is as follows. In Section
\ref{prelim}, we formalize the problem rigorously and spell out our basic
assumptions. In Section \ref{optimalw}, we characterize the optimal
correlator, $\{w_t^*\}$, for a given signal, $\{s_t\}$, subject to the power constraint.
In Section \ref{jointopt}, we address the problem of joint optimization of both $\{w_t\}$
and $\{s_t\}$, both under power constraints, and finally in Section \ref{corr+energy},
we outline extensions to wider classes of detectors that are based on correlation and energy.

\section{Assumptions and Preliminaries}
\label{prelim}

Consider the signal detection model described in the fifth paragraph of the
Introduction. We assume that $Z_1,\ldots,Z_n$ are independent copies of a
zero--mean random
variable (RV), $Z$, that has a
symmetric\footnote{The symmetry assumption is imposed mostly for convenience, but
the results can be extended to address also non-symmetric PDFs.}
PDF, $f_Z(z)$, around the origin, and
that it has a finite cumulant
generating function (CGF),
\begin{equation}
\label{cgf}
C(v)\dfn\ln\bE\{e^{vZ}\},
\end{equation}
at least in a certain interval of the real valued variable $v$. Note that since
$f_Z(\cdot)$ is assumed symmetric
around the origin, then so is $C(\cdot)$. We also assume
that $C(\cdot)$ is twice differentiable within the range it exists. It is well
known to be a convex function, because its second derivative cannot be
negative, as it can be viewed
as the variance of $Z$ under the PDF that is proportional to $f_Z(z)e^{vz}$.
Further assumptions on $Z$ and its CGF will be
spelled out in the sequel, at the places they will be needed. The following simple special cases
will accompany our derivations and discussions repeatedly in the sequel:\\

\noindent
Case 1. $Z$ is zero-mean, Gaussian RV with variance $\sigma_Z^2$:
\begin{equation}
\label{gaussian}
C(v)=\frac{\sigma_Z^2v^2}{2}.
\end{equation}

\noindent
Case 2. $Z$ is a Laplacian RV with parameter $q$, i.e.,
$f_Z(z)=\frac{q}{2}e^{-q|z|}$:
\begin{equation}
\label{laplacian}
C(v)=-\ln\left(1-\frac{v^2}{q^2}\right).
\end{equation}

\noindent
Case 3. $Z$ in a binary RV, taking values in $\{-z_0,+z_0\}$ with equal probabilities:
\begin{equation}
\label{binary}
C(v)=\ln\cosh(z_0v).
\end{equation}

\noindent
Case 4. $Z$ is a uniformly distributed RV across the interval $[-z_0,+z_0]$:
\begin{equation}
\label{uniform}
C(v)=\ln\left(\frac{\sinh(z_0v)}{z_0v}\right).
\end{equation}

The signal vector, $\bs=(s_1,\ldots,s_n)$, $s_t\in\reals$, $t=1,\ldots,n$, is assumed known, and we denote
its power by $P(\bs)$, that is,
\begin{equation}
\label{Ps}
P(\bs)\dfn\frac{1}{n}\sum_{t=1}^ns_t^2.
\end{equation}

Consider the class of correlation detectors, i.e., detectors that compare the
correlation, $\sum_{t=1}^nw_tY_t$, to a certain threshold $T$, where
$\bw=(w_1,\ldots,w_n)$ is a vector of real valued correlator coefficients, henceforth
referred to as the {\em correlator}, for short. The decision rule is as
follows: If $\sum_{t=1}^nw_tY_t < T$, accept the null hypothesis, $\calH_0$,
otherwise, accept the alternative, $\calH_1$. The threshold, $T$, controls the
trade-off between the FA probability and the MD probability of the detector.
To allow exponential decay (as $n$ grows without bound) of both types of error
probabilities, we let $T$ vary linearly with $n$, and denote $T=\theta n$,
where $\theta$ is a real valued constant, independent of $n$.

In order to have a well-defined asymptotic FA exponent, we assume that the
correlator, $\bw$, has a fixed power,
\begin{equation}
\label{Pw}
P(\bw)=\frac{1}{n}\sum_{t=1}^nw_t^2,
\end{equation}
which is independent of $n$, or, more generally, that the right--hand side (RHS) of eq.\
(\ref{Pw}) tends to a certain fixed positive power level, as $n\to\infty$ (otherwise, the
normalized logarithm of the FA probability would oscillate indefinitely, without a limit).
Indeed, the FA probability of the correlation detector is given by
\begin{equation}
\label{faprob}
P_{\mbox{\tiny FA}}=\mbox{Pr}\left\{\sum_{t=1}^nw_tN_t\ge \theta n
\right\}=Q\left(\frac{\theta n}{\sigma\|\bw\|}\right)\exe
\exp\left\{-\frac{\theta^2 n}{2\sigma_N^2 P(\bw)}\right\},
\end{equation}
where $Q$ is the well-known $Q$-function,
\begin{equation}
Q(u)\dfn\frac{1}{\sqrt{2\pi}}\int_u^{\infty}e^{-u^2/2}\mbox{d}u,
\end{equation}
and $\exe$ denotes equivalence in the exponential scale, in other words, the
notation, $a_n\exe
b_n$, for two positive sequences $\{a_n\}$ and $\{b_n\}$, means that
$\lim_{n\to\infty}\frac{1}{n}\log\frac{a_n}{b_n}=0$.
It follows from (\ref{faprob}) that the FA exponent is given by
\begin{equation} 
\label{efa}
E_{\mbox{\tiny FA}}(\theta)=\frac{\theta^2}{2\sigma_N^2P(\bw)}.
\end{equation}
Thus, the FA exponent depends on $\bw$ only via $P(\bw)$. It follows that for a
given $\theta$, if we wish to achieve a given, prescribed FA exponent,
$E_{\mbox{\tiny FA}}(\theta)\ge E_{\mbox{\tiny FA}}$ (where $E_{\mbox{\tiny
FA}}$ is a given positive number), we must have 
\begin{equation}
P(\bw)\le
P_w\dfn\frac{\theta^2}{2\sigma_N^2E_{\mbox{\tiny FA}}}. 
\end{equation}
In other words, a constraint on the
FA exponent means a corresponding constraint on the asymptotic power of $\bw$
to be no larger than $P_w$.

In order to have a well--defined MD exponent, our assumptions concerning the
asymptotic behavior of $\bw$ and $\bs$ will have to be more restrictive: We
will assume that as $n\to\infty$, the pairs $\{(w_t,s_t)\}_{t=1}^n$ obey a
certain joint PDF, $f_{WS}(w,s)$, in the following sense: For every $\lambda\ge 0$,
\begin{eqnarray}
& &\lim_{n\to\infty}\left\{\lambda\left(\cdot\frac{1}{n}w_ts_t-\theta\right)-\frac{1}{n}\sum_{t=1}^nC(\lambda w_t)-
\frac{\lambda^2\sigma_N^2}{2}\cdot\frac{1}{n}\sum_{t=1}^nw_t^2\right\}\nonumber\\
&=&\bE_{WS}\left\{
\lambda(\bE\{W\cdot S\}-\theta)-\bE\{C(\lambda
W)\}-\frac{\lambda^2\sigma_N^2}{2}\cdot\bE\{W^2\}\right\},
\end{eqnarray}
where $\bE_{WS}\{\cdot\}$ denotes
expectation with respect to (w.r.t.) $f_{WS}$. 
Whenever there is no room for
confusion, the subscript $WS$ will be omitted and the expectation 
will be denoted simply by $\bE\{\cdot\}$.
The function $f_{WS}(\cdot,\cdot)$
will be referred to as the {\em asymptotic empirical joint PDF} of $\bw$ and
$\bs$.\footnote{In the sequel, we will encounter one scenario where the asymptotic empirical
PDF will be irrelevant, but this scenario will 
be handled separately, in the original domain of vectors
of dimension $n$.}


The MD probability is now upper bounded, exponentially tightly,
by the Chernoff bound, as follows.
Denoting the Gaussian random variable, $U\dfn\sum_{t=1}^nw_tN_t$, we have
\begin{eqnarray}
P_{\mbox{\tiny MD}}&=&\mbox{Pr}\left\{\sum_{t=1}^nw_ts_t+\sum_tw_tZ_t+U\le\theta
n\right\}\nonumber\\
&\le&\inf_{\lambda\ge 0}\bE\left(\exp\left\{\lambda\left[\theta
n-\sum_{t=1}^nw_ts_t-\sum_tw_tZ_t-U\right]\right\}\right)\nonumber\\
&=&\inf_{\lambda\ge 0}\exp\left\{\lambda\left[\theta
n-\sum_{t=1}^nw_ts_t\right]\right\}\cdot\bE\exp\{-\lambda U\}\cdot
\bE\exp\left\{-\lambda\sum_{t=1}^nw_tZ_t\right\}\nonumber\\
&=&\inf_{\lambda\ge 0}\exp\left\{\lambda\left[\theta
n-\sum_{t=1}^nw_ts_t\right]\right\}\cdot\exp\left\{\frac{n\lambda^2\sigma_N^2P(\bw)}{2}\right\}\cdot\prod_{t=1}^n
\bE\exp\{-\lambda w_tZ_t\}\nonumber\\
&=&\inf_{\lambda\ge 0}\exp\left\{\lambda\left[\theta
n-\sum_{t=1}^nw_ts_t\right]\right\}\cdot\exp\left\{\frac{n\lambda^2\sigma_N^2P(\bw)}{2}\right\}\cdot\prod_{t=1}^n
\exp\{C(-\lambda w_tZ_t)\}\nonumber\\
&=&\inf_{\lambda\ge 0}\exp\left\{\lambda\left[\theta
n-\sum_{t=1}^nw_ts_t\right]\right\}\cdot\exp\left\{\frac{n\lambda^2\sigma_N^2P(\bw)}{2}\right\}\cdot\prod_{t=1}^n
\exp\{C(\lambda w_tZ_t)\},
\end{eqnarray}
where the last step is due to the symmetry of the $C(\cdot)$.
The resulting MD exponent is therefore given by
\begin{equation}
E_{\mbox{\tiny MD}}(\theta)=\sup_{\lambda\ge
0}\left\{\lambda(\bE\{W\cdot S\}-\theta)-\bE\{C(\lambda
W)\}-\frac{\lambda^2\sigma_N^2}{2}\cdot\bE\{W^2\}\right\},
\end{equation}
which is a functional of $f_{WS}$.

The problem of optimal correlator design for a given $\bs$,
is equivalent to the problem of finding a
conditional density, $f_{W|S}$, that maximizes the MD exponent subject to the
power constraint, $\bE\{W^2\}\le P_w$. The problem of joint design of both $\bw$
and $\bs$ is asymptotically equivalent to the problem of maximizing the MD
exponent over $\{f_{WS}\}$ subject to the power constraints, 
and $\bE\{W^2\}\le P_w$ and $\bE\{S^2\}\le P_s$, for some given $P_s> 0$. The first problem is relevant when the detector designer
has no control of the transmitted signal, for example, when the transmitter
and the receiver are hostile parties, which is typically the case in military
applications. The second problem is relevant when the transmitter and the
receiver cooperate. In radar applications, for example, the transmitter and the receiver
are the same party. In sections \ref{optimalw} and \ref{jointopt}, we address
the first problem and the second problem, respectively.

\noindent
{\em Comment 1.} Instead of maximizing the 
MD exponent for a fixed threshold, $\theta$, and a fixed power constraint, $P_w$, in order to fit
a prescribed FA exponent, 
in principle, there is an alternative approach: maximize the MD exponent 
directly for a given FA exponent, by
substituting $\theta =\sigma\sqrt{2P_wE_{\mbox{\tiny FA}}}$ in the MD exponent expression. 
Not surprisingly, in this case, the MD exponent would become
invariant to scaling in $W$ 
(as any scaling of $W$ can be absorbed in $\lambda$ for all terms of the MD exponent),
and so, there would be no need for the $P_w$-constraint,
but this invariance 
property holds only after maximizing over $\lambda$, not for a given $\lambda$.
However, maximizing over $\lambda$ as a first step 
of the calculation,
does not seem to lend itself to closed form analysis, in general, and consequently, 
it would make the
subsequent optimization extremely difficult, if not impossible, to carry out.
We therefore opt to fix both $\theta$ and $P_w$ throughout our derivations.

\section{Optimum Correlator for a Given Signal}
\label{optimalw}

In view of the discussion in Section \ref{prelim}, we wish to find the optimal
conditional density, $f_{W|S}$, in the sense of maximizing 
\begin{equation}
\lambda\bE\{W\cdot S\}-\bE\{C(\lambda
W)\}-\frac{\lambda^2\sigma_N^2}{2}\bE\{W^2\}=\int_{\infty}^{+\infty}f_S(s)\cdot\bE\left\{\lambda sW-C(\lambda
W)-\frac{\lambda^2\sigma_N^2}{2}\cdot W^2\bigg|S=s\right\}\mbox{d}s,
\end{equation}
subject to the power constraint,
\begin{equation}
\bE\{W^2\}\equiv\int_{\infty}^{+\infty}f_S(s)\bE\{W^2|S=s\}\mbox{d}s\le P_w.
\end{equation}
At this stage, we carry out this optimization for a given $\lambda\ge 0$, but with the
understanding that eventually, $\lambda$ will be subjected to optimization as well.
To this end, let us denote the derivative of $C(v)$ by $\dot{C}(v)$, and for a given $\rho\ge 0$, define the function
\begin{equation}
	g(w|\rho,\lambda)\dfn\dot{C}(\lambda w)+\left(\frac{\rho}{\lambda}+\sigma_N^2\lambda\right)\cdot w.
\end{equation}
Observe that since $C$ is convex, $\dot{C}$ is monotonically non-decreasing, and so, $g(\cdot|\rho,\lambda)$ is monotonically strictly
increasing, which in turn implies that it has an inverse. We denote the inverse of $g(\cdot|\rho,\lambda)$ by
$g^{-1}(\cdot|\rho,\lambda)$.
Also, since $Z$ is assumed zero mean, then $\dot{C}(0)=0$, and 
hence also $g(0|\rho,\lambda)=0$ and $g^{-1}(0|\rho,\lambda)=0$.
Note also that $g(\cdot|\rho,\lambda)$ (and hence also $g^{-1}(\cdot|\rho,\lambda)$) is a linear function if and only if $Z$ is Gaussian.
The following lemma characterizes the optimal $f_{W|S}$.

\begin{theorem}
\label{optimalWgivenS}
Let the assumptions of Section \ref{prelim} hold. Assume further that $P_w$ is such that there exists $\rho\ge 0$ (possibly,
depending on $\lambda$), with $\bE\{[g^{-1}(S|\rho,\lambda)]^2\}=P_w$. Otherwise, if $\bE\{[g^{-1}(S|0,\lambda)]^2\} < P_w$, set
$\rho=0$. Then, the optimal conditional density, $f_{W|S}$, is given by
\begin{equation}
\label{deltafunction}
f_{W|S}^*(w|s)=\delta(w-g^{-1}(s|\rho,\lambda)), 
\end{equation}
where $\delta(\cdot)$ is the Dirac delta function.
\end{theorem}

The theorem tells that the best 
correlator, $\bw^*=(w_1^*,\ldots,w_n^*)$, for a given $\bs=(s_1,\ldots,s_n)$, is obtained by
the relation,
\begin{equation}
w_t^*= g^{-1}(s_t|\rho,\lambda),~~~~~t=1,\ldots,n,
\end{equation}
which means that $w_t^*$ is given by a function of $s_t$, which is non-linear unless $Z$ is Gaussian.
To gain an initial insight regarding the condition on $\rho$, consider the Gaussian example 
(Case 1, eq.\ (\ref{gaussian})). In this case,
$g(W|\rho,\lambda)=[(\sigma_N^2+\sigma_Z^2)\lambda+\rho/\lambda]W$, and so,
$g^{-1}(S|\rho,\lambda)=\lambda S/[(\sigma_N^2+\sigma_Z^2)\lambda^2+\rho]$, whose power is $P_w$ for
$\rho=\lambda\sqrt{\bE\{S^2\}/P_w}-(\sigma_N^2+\sigma_Z^2)\lambda^2$, which is non-negative as long as 
$P_w\le \bE\{S^2\}/[(\sigma_N^2+\sigma_Z^2)^2\lambda]$. In general, the 
exact choice of $P_w$ is not crucial, as the prescribed FA exponent can be
achieved by adjusting $\theta$ proportionally 
to $\sqrt{P_w}$. However, once $P_w$ is chosen, we will keep it 
fixed throughout (see Comment 1 above).

\noindent
{\em Proof of Theorem \ref{optimalWgivenS}.}
Consider the following chain of inequalities and inequalities.
\begin{eqnarray}
& &\sup_{\{f_{W|S}:~\bE\{W^2\}\le P_w\}}
\int_{\infty}^{+\infty}f_S(s)\cdot\bE\left\{\lambda sW-C(\lambda
W)-\frac{\lambda^2\sigma_N^2}{2}\cdot W^2\bigg|S=s\right\}\mbox{d}s\nonumber\\
&=&\sup_{f_{W|S}}\inf_{\varrho\ge 0}\bigg\{
\int_{\infty}^{+\infty}f_S(s)\cdot\bE\left\{\lambda sW-C(\lambda
W)-\frac{\lambda^2\sigma_N^2}{2}\cdot W^2\bigg|S=s\right\}\mbox{d}s+\nonumber\\
& &\frac{\varrho}{2}\left[P_w-
\int_{\infty}^{+\infty}f_S(s)\bE\{W^2|S=s\}\mbox{d}s\right]\bigg\}\nonumber\\
&\eqa&\inf_{\varrho\ge 0}\sup_{f_{W|S}}
\bigg\{\int_{\infty}^{+\infty}f_S(s)\cdot\bE\left\{\lambda sW-C(\lambda
W)-\left(\frac{\lambda^2\sigma_N^2}{2}+\frac{\varrho}{2}\right)\cdot W^2\bigg|S=s\right\}\mbox{d}s+\frac{\varrho P_w}{2}\bigg\}\nonumber\\
&\eqb&\inf_{\varrho\ge 0}
\left\{\int_{\infty}^{+\infty}f_S(s)\cdot\sup_w\left\{\lambda sw-C(\lambda
w)-\left(\frac{\lambda^2\sigma_N^2}{2}+\frac{\varrho}{2}\right)\cdot w^2\right\}\mbox{d}s+\frac{\varrho P_w}{2}\right\}\nonumber\\
&\eqc&\inf_{\varrho\ge 0}
\left\{\int_{\infty}^{+\infty}f_S(s)\cdot\left\{\lambda sg^{-1}(s|\varrho,\lambda)-C(\lambda
g^{-1}(s|\varrho,\lambda))-\left(\frac{\lambda^2\sigma_N^2}{2}+
\frac{\varrho}{2}\right)\cdot [g^{-1}(s|\varrho,\lambda)]^2\right\}\mbox{d}s+\frac{\varrho P_w}{2}\right\}\nonumber\\
&=&\inf_{\varrho\ge 0}
\bigg\{\int_{\infty}^{+\infty}f_S(s)\cdot\left\{\lambda sg^{-1}(s|\varrho,\lambda)-C(\lambda
g^{-1}(s|\varrho,\lambda))-\frac{\lambda^2\sigma_N^2}{2}
\cdot [g^{-1}(s|\varrho,\lambda)]^2\right\}\mbox{d}s+\nonumber\\
& &\frac{\varrho}{2}\left[P_w-\int_{\infty}^{+\infty}f_S(s)\cdot[g^{-1}(s|\varrho,\lambda)]^2\mbox{d}s\right]\bigg\}\nonumber\\
&\led&\int_{\infty}^{+\infty}f_S(s)\cdot\left\{\lambda sg^{-1}(s|\rho,\lambda)-C(\lambda
g^{-1}(s|\rho,\lambda))-\frac{\lambda^2\sigma_N^2}{2}
\cdot [g^{-1}(s|\rho,\lambda)]^2\right\}\mbox{d}s+\nonumber\\
& &\frac{\rho}{2}\left[P_w-\int_{\infty}^{+\infty}f_S(s)\cdot[g^{-1}(s|\rho,\lambda)]^2\mbox{d}s\right]\nonumber\\
&\eqe&\int_{\infty}^{+\infty}f_S(s)\cdot\left\{\lambda sg^{-1}(s|\rho,\lambda)-C(\lambda
g^{-1}(s|\rho,\lambda))-\frac{\lambda^2\sigma_N^2}{2}
\cdot [g^{-1}(s|\rho,\lambda)]^2\right\}\mbox{d}s\nonumber\\
&=&\bE\left\{\lambda Sg^{-1}(S|\rho,\lambda)-C(\lambda
g^{-1}(S|\rho,\lambda))-\frac{\lambda^2\sigma_N^2}{2}
\cdot [g^{-1}(S|\rho,\lambda)]^2\right\},
\end{eqnarray}
where (a) is since the objective is affine in both $f_{W|S}$ and in $\rho$ (and hence is concave in $f_{W|S}$ and
convex in $\varrho$), (b) is since the unconstrained maximum of the conditional expectation 
of a function of $W$ given $S=s$ is attained
when $f_{W|S}$ puts all its mass on the maximum of this function, (c)
is because the maximum is over a concave function of $w$, which is attained at the point of zero-derivative,
$w=g^{-1}(s|\varrho,\lambda)$, (d) is by the postulate that $\rho\ge 0$, and (e) is because either $\rho=0$ or
$P_w-\bE\{[g^{-1}(S|\rho,\lambda)]^2\}=0$. The upper bound on the constrained maximum in the first line of the
above chain is therefore 
attained by $W=g^{-1}(S|\rho,\lambda)$ with probability one, which is equivalent to (\ref{deltafunction}).
This completes the proof of Theorem \ref{optimalWgivenS}. $\Box$\\

\noindent
{\bf Optimal correlator weights within a finite set.}
There is a practical motivation to consider the case where 
$\bw=(w_1,\ldots,w_n)$ is restricted to be a binary vector with bipolar components,
taking the values $+\sqrt{P_w}$ and $-\sqrt{P_w}$ only. The reason is that in such a case, 
the implementation of the correlation detector involves no
multiplications at all, as it is equivalent to the comparison of the difference 
$$\sum_{\{t:~w_t=\sqrt{P_w}\}}Y_t-
\sum_{\{t:~w_t=-\sqrt{P_w}\}}Y_t$$ 
to $\theta n/\sqrt{P_w}$.
Here, the maximization over $w$ (step (b) in the proof of Theorem \ref{optimalWgivenS})
is carried out just over its two allowed values, $+\sqrt{P_w}$ and
$-\sqrt{P_w}$. As
$C(\cdot)$ is symmetric, the maximum is readily seen to be attained by
$W=\sqrt{P_w}\cdot\mbox{sgn}(S)$, which means
$w_t^*=\sqrt{P_w}\cdot\mbox{sgn}(s_t)$. 

Suppose, more generally, that $\{w_t\}$ is constrained to take on values in a finite set
whose cardinality $k$ is fixed, independent of $n$. 
This can be considered as a compromise between the above two extremes of performance and computational complexity,
since the number of multiplications need not be larger than $k-1$.
The design of such a signal is very similar to the scalar quantizer design problem: 
A finite--alphabet
signal $w_t$ is defined as follows. Let $s_{\min}\equiv a_0 < a_1 < \ldots <
a_{k-1} < a_k\equiv s_{\max}$, where $s_{\min}=\min_ts_t$, $s_{\max}=\max_ts_t$,
and let $\calI_i\dfn [a_i, a_{i+1})$,
$i=0,1,\ldots,k-1$, be given. Define
\begin{equation}
W=\sum_{i=0}^{k-1}\omega_i\cdot 1\{S\in\calI_i\},
\end{equation}
for some given $\omega_0,\omega_1,\ldots,\omega_{k-1}$.
We wish to minimize
\begin{equation}
\Delta=\sum_{i=0}^{k-1}\int_{a_i}^{a_{i+1}}
\mbox{d}s\cdot
	f_S(s)\left[\lambda\omega_is-C(\lambda\omega_i)-\frac{1}{2}\lambda^2\sigma_N^2\omega_i^2+\frac{\rho}{2}(P_w-\omega_i^2)\right]
\end{equation}
over $\{a_i\}$, $i=1,\ldots,k-1$, and $\{\omega_i\}$,
$i=0,1,\ldots,k-1$.
The necessary conditions for optimality are obtained by equating to zero all partial
derivatives w.r.t.\ $\{a_i\}$, $i=1,\ldots,k-1$, and $\{\omega_i\}$,
$i=0,1,\ldots,k-1$. This results in the following sets of equations:
\begin{eqnarray}
\lambda\omega_{i-1}a_i-C(\lambda\omega_{i-1})-\left(\frac{\rho}{2}+\frac{\lambda^2\sigma_N^2}{2}\right)\omega_{i-1}^2&=&
\lambda\omega_ia_i-C(\lambda\omega_i)-
	\left(\frac{\rho}{2}+\frac{\lambda^2\sigma_N^2}{2}\right)\omega_i^2,~~i=1,2,\ldots,k-1\nonumber\\
	\dot{C}(\lambda\omega_i)+\left(\frac{\rho}{\lambda}+
	\lambda\sigma_N^2\right)\omega_i&=&\bE\{S|S\in\calI_i\},~~i=0,1,\ldots,k-1.\nonumber
\end{eqnarray}
Alternatively, we may represent these equations as:
\begin{eqnarray}
	a_i&=&\frac{C(\lambda\omega_i)-C(\lambda\omega_{i-1})+(\rho+\lambda^2\sigma_N^2)
	(\omega_i^2-\omega_{i-1}^2)/2}{\lambda(\omega_i-\omega_{i-1})},~~~~i=1,2,\ldots,k-1\\
	\omega_i&=&g^{-1}[\bE\{S|S\in\calI_i\}|\rho,\lambda\}],~~~~i=0,1,\ldots,k-1,
\end{eqnarray}
where $\rho$ is tuned such that
\begin{equation}
\sum_{i=0}^{k-1}P(\calI_i)\cdot
(g^{-1}[\bE\{S|S\in\calI_i\}|\rho,\lambda\}])^2\le P_w
\end{equation}
as before.
The first set of equations is parallel to the nearest-neighbor condition in
optimal quantizer design, and the second set corresponds to the centroid
condition. Optimal signal design can be conducted iteratively, like in
quantizer design, by alternating between the two sets of equations.

\noindent
{\em Example 1.}
Consider the case where $Z\sim\calN(0,\sigma_Z^2)$ (i.e., Case 1). In this
case, $C(v)=\sigma_Z^2v^2/2$, and so, $\dot{C}(v)=\sigma_Z^2v$, which leads to
\begin{equation}
	g(w|\rho,\lambda)=\sigma_Z^2\lambda w+\left(\frac{\rho}{\lambda}+\sigma_N^2\lambda\right)\cdot
	w=\left[(\sigma_N^2+\sigma_Z^2)\lambda+\frac{\rho}{\lambda}\right]\cdot w,
\end{equation}
and so,
\begin{equation}
g^{-1}(s|\rho,\lambda)=\frac{\lambda
	s}{(\sigma_N^2+\sigma_Z^2)\lambda^2+\rho}.
\end{equation}
Choosing
\begin{equation}
	\rho=\lambda\sqrt{\frac{\bE\{S^2\}}{P_w}}-\lambda^2\sigma_Z^2,
\end{equation}
yields
\begin{equation}
	w_t^*=\sqrt{\frac{P_w}{\bE\{S^2\}}}\cdot s_t,
\end{equation}
which results in
\begin{equation}
E_{\mbox{\tiny MD}}(\theta)=\left\{\begin{array}{ll}
\frac{(\sqrt{P_w
	\bE\{S^2\}}-\theta)^2}{2(\sigma_N^2+\sigma_Z^2)P_w} & \theta < \sqrt{P_w \bE\{S^2\}}\\
0 & \theta \ge \sqrt{P_w \bE\{S^2\}}\end{array}\right.
\end{equation}
If $w_t$ is constrained to be binary, then
as we already saw, $w_t^*=\sqrt{P_w}\cdot\mbox{sgn}(s_t)$ and then
\begin{equation}
E_{\mbox{\tiny MD}}(\theta)=\left\{\begin{array}{ll}
	\frac{(\sqrt{P_w}\cdot\bE\{|S|\}
	-\theta)^2}{2(\sigma_N^2+\sigma_Z^2)P_w} & \theta < \sqrt{P_w}\cdot\bE\{|S|\}\\
0 & \theta \ge \sqrt{P_w}\cdot\bE\{|S|\}\end{array}\right.
\end{equation}
For the more general quantization, we obtain
\begin{equation}
a_i=\frac{(\sigma_Z^2\lambda^2/2+\rho/2)(\omega_i^2-\omega_{i-1}^2)}
        {\lambda(\omega_i-\omega_{i-1})}=\left(\sigma_Z^2\lambda+\frac{\rho}{\lambda}
\right)\cdot\frac{\omega_i+\omega_{i-1}}{2}.
\end{equation}
For simplicity, let us assume that $f_S$ is uniform across
the interval $[-A,+A]$. In this case, $\bE\{S|S\in\calI_i\}=(a_i+a_{i+1})/2$,
and so,
\begin{equation}
	\omega_i=\frac{\lambda(a_i+a_{i+1})}{2[(\sigma_N^2+\sigma_Z^2)\lambda^2+\rho)]}.
\end{equation}
It follows that $\{a_i\}$ have uniform spacings across the support of $S$,
that is, $a_i=(2i/k-1)S$, $i=0,1,\ldots,k$. Accordingly,
\begin{equation}
	\omega_i=\frac{\lambda A[(2i+1)/k-1]}{(\sigma_N^2+\sigma_Z^2)\lambda^2+\rho},
\end{equation}
where $\rho$ is chosen such that
\begin{equation}
	\frac{1}{k}\sum_{i=0}^{k-1}\frac{\lambda^2A^2[(2i+1)/k-1]^2}{[(\sigma_N^2+\sigma_Z^2)\lambda^2+\rho]^2}=P_w.
\end{equation}
The binary case considered above is the special case of $k=2$.
This concludes Example 1. $\Box$

If $\{s_t\}$ is itself a finite--alphabet signal, then the optimal $\{w_t^*\}$
is also a finite-alphabet signal with the same alphabet size, even without restricting it to be so in the first place. If this alphabet
is small enough and/or there is a strong degree of symmetry, one might as well
optimize the levels of $\{w_t\}$ directly subject to the power constraint.
Consider, for example, the case of a 4-ASK signal, $s_t\in\{-3a,-a,+a,+3a\}$, for some given
$a>0$. Then, since the PDF of $Z$
is assumed symmetric, the alphabet of the optimal
$\{w_t\}$ must be of the form $\{-\beta,-\alpha,+\alpha,+\beta\}$ for some
$0< \alpha < \beta$. Assuming that $s_t=\pm a$ along half of the time and
$s_t=\pm 3a$ along the other half, then
$w_t=\pm \alpha$ and
$w_t=\pm \beta$ in the corresponding halves, and so, $\frac{1}{2}\alpha^2+\frac{1}{2}\beta^2=P_w$, or
$\beta=\sqrt{2P_w-\alpha^2}$. Thus, the MD exponent should be maximized over
one parameter only (beyond the optimization over $\lambda$), which is $\alpha\in[0,\sqrt{2P_w}]$. In particular,
\begin{eqnarray}
\label{4ask}
E_{\mbox{\tiny MD}}(\theta)&=&\sup_{\lambda\ge
0}\max_{0\le\alpha\le\sqrt{2P_w}}\bigg\{\frac{1}{2}\lambda a\alpha
+\frac{3}{2}\lambda a\sqrt{2P_w-\alpha^2} -\nonumber\\
& &\frac{1}{2}C(\lambda\alpha)-\frac{1}{2}C(\lambda\sqrt{2P_w-\alpha^2})
-\lambda\theta-\frac{\lambda^2\sigma_N^2P_w}{2}\bigg\}.
\end{eqnarray}

We next examine this expression in several examples.

\begin{figure}[h!t!b!]
\centering
\includegraphics[width=8.5cm, height=8.5cm]{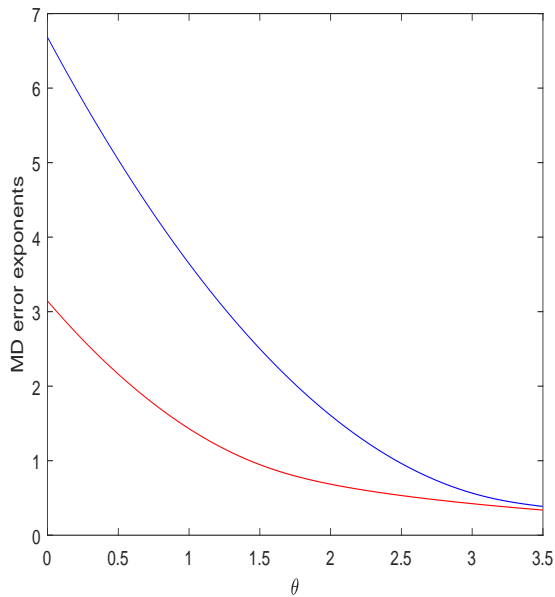}
\caption{Graphs for binary interference: MD error exponents as functions of $\theta$ pertaining to the classical correlator
(red curve) and the optimal correlator (blue curve) for the following parameter
values: $P_w=1$, $z_0=7$, $a=4$, and $\sigma_N^2=1$.}
\label{graph1}
\end{figure}

\noindent
{\em Example 2.} 
Let $Z$ be a binary symmetric source, taking values $\pm z_0$ for some $z_0> 0$ (Case 3).
Then, owing to eq.\ (\ref{binary}), 
eq.\ (\ref{4ask}) becomes
\begin{eqnarray}
E_{\mbox{\tiny MD}}(\theta)&=&\sup_{\lambda\ge
0}\max_{0\le\alpha\le\sqrt{2P_w}}\bigg\{\frac{1}{2}\lambda a\alpha
+\frac{3}{2}\lambda a\sqrt{2P_w-\alpha^2} -\nonumber\\
& &\frac{1}{2}\ln\cosh(z_0\lambda\alpha)-\frac{1}{2}\ln\cosh(z_0\lambda\sqrt{2P_w-\alpha^2})
-\lambda\theta-\frac{\lambda^2\sigma_N^2P_w}{2}\bigg\}\\
&=&\frac{1}{2}\sup_{\lambda\ge
0}\max_{0\le\alpha\le\sqrt{2P_w}}\bigg\{\lambda a\alpha
+3\lambda a\sqrt{2P_w-\alpha^2} -\nonumber\\
& &\ln\cosh(z_0\lambda\alpha)-\ln\cosh(z_0\lambda\sqrt{2P_w-\alpha^2})
-2\lambda\theta-\lambda^2\sigma_N^2P_w\bigg\}
\end{eqnarray}
The `classical' correlator, where $w_t\propto s_t$, corresponds to the choice
$\alpha=\sqrt{P_w/5}$ instead of maximizing over $\alpha$. In Fig.\ \ref{graph1}, we compare the
two curves of the MD exponent as functions of $\theta$. Since they share the same level of $P_w$,
the FA exponents are the same for a given $\theta$. As can be seen, the optimal correlator significantly outperforms the
classical one, which is optimal in the Gaussian case only.
This concludes Example 2. $\Box$

\begin{figure}[h!t!b!]
\centering
\includegraphics[width=8.5cm, height=8.5cm]{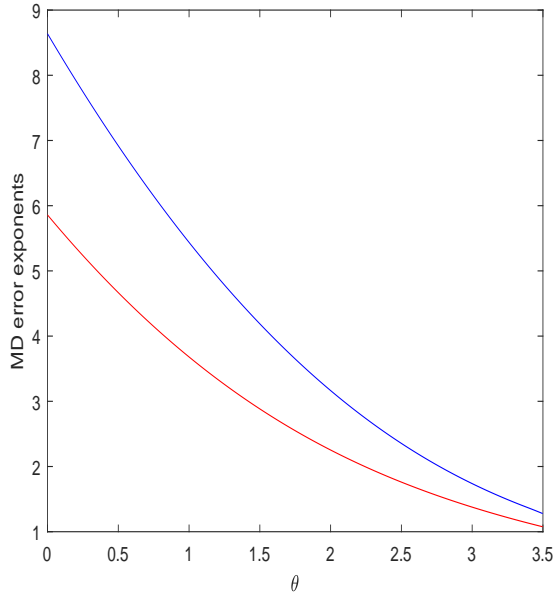}
\caption{Graphs for uniformly distributed interference: MD error exponents as functions of $\theta$ pertaining to the classical
correlator (red
curve) and the optimal correlator (blue curve) for the following parameter
values: $P_w=1$, $z_0=7$, $a=4$, and $\sigma_N^2=1$.}
\label{graph2}
\end{figure}

\noindent
{\em Example 3.} 
We conduct a similar comparison for the case
where $Z$ is distributed uniformly over $[-z_0,+z_0]$ (Case 4),
which corresponds to eq.\ (\ref{uniform}).
The results are displayed in Fig.\ \ref{graph2}, and as can be seen, here too,
the optimal correlator significantly improves upon the classical one.
This concludes Example 3. $\Box$

It is interesting to note that in both Examples 2 and 3, for large $\theta$, the two graphs approach each other
faster than they approach zero. A possible intuitive explanation is that for large
$\theta$, what counts is the behavior of the PDF of $\sum_tw_tZ_t$, fairly close to
its peak, where the regime of the central limit theorem is quite relevant, and so, there is no significant
difference from Case 1, where $Z$ is Gaussian and the classical correlator
is good. Mathematically, as $\theta$ grows, the optimum $\lambda$ decreases, and
so, it `samples' the function $C(\lambda w_t)$ in the vicinity of the origin, where it is
well approximated by a quadratic function, just like in the Gaussian case (Case 1).\\

\begin{figure}[h!t!b!]
\centering
\includegraphics[width=8.5cm, height=8.5cm]{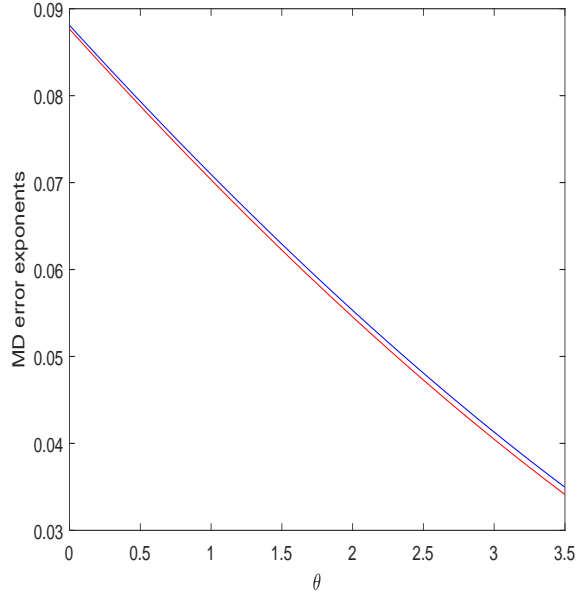}
\caption{Graphs for Laplace-distributed interference: MD error exponents as
functions of $\theta$ pertaining to the classical
correlator (red
curve) and the optimal correlator (blue curve) for the following parameter
values: $P_w=1$, $q=0.1$, $a=4$, and $\sigma_N^2=1$.}
\label{graph3}
\end{figure}

\noindent
{\em Example 4.}
Finally, consider the case where $Z$ is Laplacian (Case 2).
In this case, the differences turned out to be rather minor -- see Fig.\ \ref{graph3}. A plausible intuition is
that the Laplacian PDF is much `closer' to the Gaussian PDF, relative to the
binary distribution and the uniform distribution of Examples 2 and 3.
This concludes Example 4. $\Box$

The loss relative to the optimal LRT detector depends on the relative intensity of the
process $\{Z_t\}$ compared to the Gaussian noise component.

\section{Joint Optimization of the Correlator and the Signal}
\label{jointopt}

So far, we have concerned ourselves with the optimization of the correlator
waveform, $\{w_t\}$ for a given signal, $\{s_t\}$. But what would be the
optimal signal $\{s_t\}$ (subject to a power constraint) when it is jointly optimized with $\{w_t\}$?
Mathematically, we are interested in the problem,
\begin{eqnarray}
\label{main}
& &\sup_{\{f_S:~\bE\{S^2\}\le P_s\}}
\sup_{\{f_{W|S}:~\bE\{W^2\}\le P_w\}}
E_{\mbox{\tiny MD}}(\theta)\nonumber\\
&=&\sup_{\{f_S:~\bE\{S^2\}\le P_s\}}
\sup_{\{f_{W|S}:~\bE\{W^2\}\le P_w\}}
\sup_{\lambda\ge 0}\left[\bE\left\{\lambda (W\cdot S-\theta)-C(\lambda
W)-\frac{\lambda^2\sigma_N^2W^2}{2}\right\}\right]\nonumber\\
&=&\sup_{\{f_W:~\bE\{W^2\}\le P_w\}}
\sup_{\lambda\ge 0}
\sup_{\{f_{S|W}:~\bE\{S^2\}\le P_s\}}
\left[\lambda\bE\{W\cdot S\}-\bE\{C(\lambda
W)\}-\lambda\theta-\frac{\lambda^2\sigma_N^2\bE\{W^2\}}{2}\right]\nonumber\\
&\eqa&\sup_{\{f_W:~\bE\{W^2\}\le P_w\}}
\sup_{\lambda\ge 0}\left[\lambda\bE\left\{
W\cdot\sqrt{\frac{P_s}{\bE\{W^2\}}}\cdot W\right\}-\bE\{C(\lambda
W)\}-\lambda\theta-\frac{\lambda^2\sigma_N^2\bE\{W^2\}}{2}\right\}\nonumber\\
&=&\sup_{\{f_W:~\bE\{W^2\}\le P_w\}}
\sup_{\lambda\ge 0}\left\{\lambda\sqrt{P_s\bE\{W^2\}}-
\bE\{C(\lambda W)\}
-\lambda\theta-\frac{\lambda^2\sigma_N^2\bE\{W^2\}}{2}\right\}\nonumber\\
&=&\sup_{\lambda\ge
0}\sup_{P\le P_w}\left\{\lambda\sqrt{P_sP}-\min_{\{f_W:~\bE\{W^2\}=P\}}
\bE\{C(\lambda W)\}
-\lambda\theta-\frac{\lambda^2\sigma_N^2P}{2}\right\},
\end{eqnarray}
where in (a) we have used the simple fact that, for a given $W$ and $P_s$, the 
correlation, $\bE\{W\cdot S\}$ is maximized by $S=\sqrt{P_s/\bE\{W^2\}}\cdot W$.
Earlier, we maximized the
MD exponent w.r.t.\ $W$ for a given $S$ and found that the optimal $W$
is given by a function,
$g^{-1}(S|\rho,\lambda)$, which is, in general, non--linear (unless $Z$ is
Gaussian). Now,
on the other hand, the optimal $S$ for a given $W$ turns out to be
given by a linear function. 
These two findings settle together if and only if 
$W$ takes values only in the set of solutions, $\calS(\zeta)$, to the equation
\begin{equation}
\dot{C}(\lambda W)+\frac{\rho}{\lambda}\cdot W=\zeta\cdot W,
\end{equation}
for some $\zeta > 0$ (and then $S$ takes the corresponding values according to their relationship). 
The two sides of the equation represent the non-linear and the linear relations, respectively.
Note that $\calS(\zeta)$
always includes at least the solution $W=0$.
Once $\zeta$ is chosen,
$W$ is allowed to take on values only within
$\calS(\zeta)$. The inner minimization over $f_W$ in the last line of
(\ref{main}) is obviously lower bounded by $\tilde{C}_\lambda(P)$, which is defined as
\begin{equation}
\label{Ctilde}
\tilde{C}_\lambda(P)\dfn\inf_{\zeta>
0}\inf_{\{\mu(\cdot):~\int_{\calS^2(\zeta)}p\cdot
\mu(p)\mbox{d}p=P\}}\int_{\calS^2(\zeta)}\mu(p)C(\lambda\sqrt{p})\mbox{d}p,
\end{equation}
where $\calS^2(\zeta)=\{w^2:~w\in\calS(\zeta)\}$,
and $\mu(\cdot)$ is understood to be a weight function over $\calS^2(\zeta)$, i.e.,
$\mu(p)\ge 0$ for all $p\in\calS^2(\zeta)$ and
$\int_{\calS^2(\zeta)}\mu(p)\mbox{d}p=1$. 
While this expression appears complicated, there are two facts that help to
simplify it significantly. The first is that,
in most cases, $\calS(\zeta)$ is a finite set (unless $C(\cdot)$ 
is linear, or contains linear segments), and the second is that
only two members of $\calS(\zeta)$ suffice, i.e., eq.\ (\ref{Ctilde})
simplifies to
\begin{equation}
\tilde{C}_\lambda(P)\dfn\inf_{\zeta>
0}\min_{\{p_0,p_1\in\calS^2(\zeta),~\alpha\in[0,1]:~(1-\alpha)p_0+\alpha
p_1=P\}}
\{(1-\alpha)C(\lambda\sqrt{p_0})+
\alpha C(\lambda\sqrt{p_1})\}.
\end{equation}
The function
$\tilde{C}_\lambda(P)$ has the flavor of a lower convex envelope for
the function $C(\lambda\sqrt{\cdot})$, but with the exception that the support of the convex combinations is
limited to $\calS^2(\zeta)$.
Finally, the optimal MD exponent is given by
\begin{equation}
	\label{MDexp}
E_{\mbox{\tiny MD}}(\theta)=\sup_{\lambda\ge
	0}\sup_{P\le P_w}\left\{\lambda(\sqrt{P_sP}-\theta)-\tilde{C}_\lambda(P)-
\frac{\lambda^2\sigma_N^2P}{2}\right\}.
\end{equation}
The optimal $W$ is one that achieves $\tilde{C}_\lambda(P)$ for the
maximizing $\lambda$ and $P$, that is, the components of $\{|w_t|\}$ take only two values 
in $\calS(\zeta^*)$, with relative frequencies given by $\alpha^*$ and
$1-\alpha^*$, where $\zeta^*$ and $\alpha^*$ are the achievers of $\tilde{C}_\lambda(P^*)$,
$P^*$ being the optimal $P$.
In other words, the optimal signal has at most four levels, $\pm
a$ and $\pm b$, for some $a\ge 0$ and $b>0$. 

\noindent
{\em Comment 2.} By a simple change of variables, $q=\lambda^2p$, it is readily seen 
that $\tilde{C}_\lambda(P)$ depends on $\lambda$ and $P$ only via the quantity 
$\lambda\sqrt{P}$, and so, it might as well be denoted as
$\tilde{C}(\lambda\sqrt{P})$. $\Box$

Observe that while the function $C(\cdot)$ is always convex, nothing general can be
asserted regarding convexity or concavity properties of the function
$C(\lambda\sqrt{\cdot})$, as the internal square root function, which is
concave, may or may not
destroy the convexity of the composite function, depending on the function $C(\cdot)$. 
In other words, $C(\lambda\sqrt{\cdot})$
may either be convex, or concave, or neither.
For example, if $Z\in\{-z_0,+z_0\}$ with equal
probabilities (as in Case 3), then
$C(\lambda\sqrt{p})=\ln\cosh(z_0\lambda\sqrt{p})$ which is concave in $p$. On the other
hand, if $Z$ is Laplacian with parameter $q$ (Case 2), then
$C(\lambda\sqrt{p})=-\ln(1-\lambda^2p/q^2)$, which is convex in $p$. By mixing these two distributions, we
can also make it neither convex, nor concave, as will be shown in the sequel.

Let us examine now several special cases, where the form of $\tilde{C}_\lambda(P)$
can be determined more explicitly.

\noindent
1.~Consider first the Gaussian case (Case 1), where
$C(\lambda\sqrt{p})=\frac{1}{2}\sigma_Z^2\lambda^2 p$, namely, it is
linear in $p$. In this case, for $\zeta=\sigma_Z^2\lambda+\rho/\lambda$,
$\calS(\zeta)=\reals^+$, the choice of $\mu$ is immaterial, and
$\tilde{C}_\lambda(P)=\frac{1}{2}\sigma_Z^2\lambda^2P$.
In this case, any signal $\bw$ with power $P$ 
is equally good, as expected.

\noindent
2.~Consider next the case where $C(\lambda\sqrt{\cdot})$ is
is convex. Then,
\begin{eqnarray}
\bE\{C(\lambda
	W)\}&=&\bE\left\{C\left(\lambda\sqrt{W^2}\right)\right\}\\
&\ge&C\left(\lambda\sqrt{\bE\{W^2\}}\right)\\
&=&C(\lambda\sqrt{P}),
\end{eqnarray}
where the inequality is achieved with equality whenever $W^2=\mbox{const}$
with probability one, and then this constant must be $P$. So, here
$\tilde{C}_\lambda(P)=C(\lambda\sqrt{P})$,
$\calS^2(\zeta)=\{0,P\}$ and $\mu(p)=\delta(p-P)$,
which is expected, because in the convex case, there is no need for any
non-trivial convex combinations. 
The optimal
signal vector $\bw$ is any member of $\{-\sqrt{P^*},+\sqrt{P^*}\}^n$, and then
$\bs$ is the corresponding member of $\{-\sqrt{P_s},+\sqrt{P_s}\}^n$.
It is interesting to note that they both turn out to be DC or bipolar signals,
which is good news from the practical point of view, as discussed in Section \ref{optimalWgivenS}.

\noindent
3.~We now move on to the case where $C(\lambda\sqrt{\cdot})$ is concave. 
In this case, it is instructive to return temporarily to the original 
domain of vectors $\{\bw\}$ of finite dimension $n$, 
find the optimal solution in that domain,
and finally, take the limit of large $n$ (see footnote no.\ 3).
We therefore wish to 
minimize $\frac{1}{n}\sum_{t=1}^nC(\lambda w_t)$ s.t.\ $\sum_{t=1}^nw_t^2=nP$.
Since $C(\lambda\sqrt{0})=0$ and each $w_t^2$ is limited to the range $[0,nP]$, we can lower bound the
function $C(\lambda\sqrt{w_t^2})$ (which is concave as a function of $w_t^2$),
by a linear function of $w_t^2$, as follows:
\begin{equation}
C\left(\lambda\sqrt{w_t^2}\right)\ge\frac{C(\lambda\sqrt{nP})}{nP}\cdot w_t^2,
\end{equation}
with equality at $w_t^2=0$ and $w_t^2=nP$. 
Consequently,
\begin{eqnarray}
\frac{1}{n}\sum_{t=1}^nC(\lambda w_t)&=&
\frac{1}{n}\sum_{t=1}^nC\left(\lambda\sqrt{w_t^2}\right)\\
&\ge&\frac{1}{n}\sum_{t=1}^n\frac{C(\lambda\sqrt{nP})}{nP}\cdot
w_t^2\\
&=&\frac{C(\lambda\sqrt{nP})}{nP}\cdot P\\
&=&\frac{C(\lambda\sqrt{nP})}{n},
\end{eqnarray}
with equality if one of the components of $\bw$ is equal to $\pm\sqrt{nP}$ and all
other components vanish, and then, the same component of $\bs$ is
$\pm\sqrt{nP_s}$ (and, of course, all other vanish), correspondingly. Here, we
have $\calS^2(\zeta)=\{0,nP\}$ and
$\mu(p)=\left(1-\frac{1}{n}\right)\delta(p)+\frac{1}{n}\delta(p-nP)$.
Asymptotically, as $n$ grows without bound,
$\tilde{C}_\lambda(P)=\lim_{n\to\infty}C(\lambda\sqrt{nP})/n$, and the limit exists since
$C(\lambda\sqrt{nP})/n$ is monotonically non-increasing by the assumed concavity of $C(\lambda\sqrt{\cdot})$.
If this limit
happens to vanish (like in Case 3, for instance), then the interference
$\{Z_t\}$ has no impact whatsoever on the MD exponent for the optimal $\bs$ and $\bw$.
Here too, the optimal signaling is binary.

We now summarize our findings, in this section so far, in the following theorem.
\begin{theorem}
Let the assumptions of Section \ref{prelim} hold. Then, $w_t^*$ and $s_t^*$ are proportional to each other
with $|w_t^*|$ and $|s_t^*|$ taking values in a finite set of size at most two ($t=1,\ldots,n$), and the MD exponent is given by
eq.\ (\ref{MDexp}).
\begin{enumerate}
\item If the function $C(\lambda\sqrt{\cdot})$ is convex, then both $\bw^*$ and $\bs^*$ are either DC or bipolar, and 
the MD exponent is given by
\begin{equation}
E_{\mbox{\tiny MD}}(\theta)=\sup_{\lambda\ge
0}\sup_{P\le P_w}\left\{\lambda(\sqrt{P_sP}-\theta)-C\left(\lambda\sqrt{P}\right)-
\frac{\lambda^2\sigma_N^2P}{2}\right\}.
\end{equation}
\item If the function $C(\lambda\sqrt{\cdot})$ is concave, then the components of both $\bw^*$ and $\bs^*$ 
are all zero, except for one component which exploits their entire energy. The
MD exponent is given by
\begin{equation}
E_{\mbox{\tiny MD}}(\theta)=\sup_{\lambda\ge
0}\sup_{P\le P_w}\left\{\lambda(\sqrt{P_sP}-\theta)-\lim_{n\to\infty}\frac{C\left(\lambda\sqrt{Pn}\right)}{n}-
\frac{\lambda^2\sigma_N^2P}{2}\right\}.
\end{equation}
\end{enumerate}
\end{theorem}

\begin{figure}[h!t!b!]
\centering
\includegraphics[width=8.5cm, height=8.5cm]{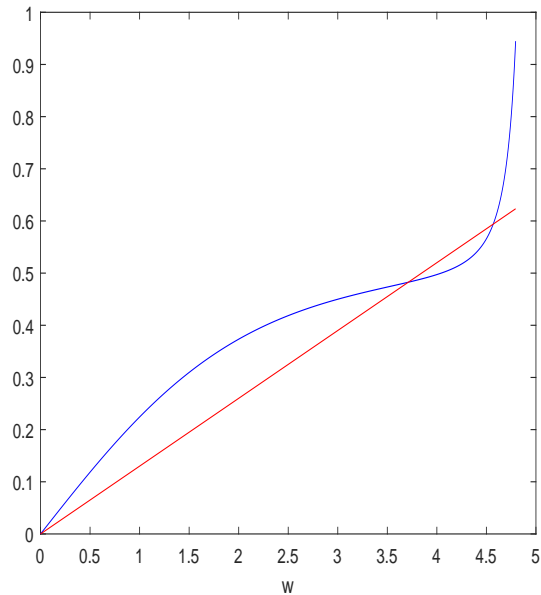}
\caption{The functions $\dot{C}(\lambda w)$ (blue curve) and $(\zeta-\rho/\lambda)\cdot w$
(red straight line) for the example of eqs.\ (\ref{nonconvexnonconcave}) and
(\ref{cdot}) with
parameter values: $\delta=0.95$, $q=5$, $z_0=0.5$, $\lambda=1$, and
$\zeta-\rho/\lambda=0.13$. As can be seen, these two graphs meet at three
points, $w_0=0$, $w_1\approx 3.71$ and $w_2\approx 4.58$.}
\label{graph4}
\end{figure}

Finally, we should consider the case where $C(\lambda\sqrt{\cdot})$ is neither convex, nor concave.
Here, we will not carry out the full calculations needed, but we will
demonstrate that $\calS(\zeta)$ may include more than one positive solution,
in addition to the trivial solution at the origin.
Consider, for example a mixture of the binary PDF and the Laplacian PDF with
weights $\delta$ and $1-\delta$, respectively ($\delta\in(0,1)$).
In this case,
\begin{equation}
\label{nonconvexnonconcave}
C(\lambda w)=C\left(\lambda\sqrt{w^2}\right)=\ln\left[\delta\cdot\cosh\left(z_0\lambda
\sqrt{w^2}\right)+\frac{1-\delta}{1-\lambda^2w^2/q^2}\right],
\end{equation}
If $\delta$ is close to 1, the hyperbolic cosine term is dominant for small and
moderate values of $w$, where $C(\lambda(\sqrt{\cdot})$ is concave. In contrast, when
$w^2$ approaches $q^2/\lambda^2$, the second term tends steeply to infinity
and hence must be convex. So in this example, $\hat{C}$ is concave in a certain
range of relatively small $w^2$ and at some point it becomes convex.
Now, the derivative w.r.t.\ $w$ is given by
\begin{equation}
\label{cdot}
\dot{C}(\lambda w)=\frac{\delta z_0\sinh(z_0\lambda
w)+(1-\delta)\cdot2\lambda
q^2/(q^2-\lambda^2w^2)^2}{\delta\cosh(z_0\lambda
w)+(1-\delta)q^2/(q^2-\lambda^2w^2)}.
\end{equation}
As discussed above, the first step is to solve the equation
\begin{equation}
\dot{C}(\lambda w)=\left(\zeta-\frac{\rho}{\lambda}\right)w.
\end{equation}
As there is no apparent analytical closed-form solution to this equation,
here we demonstrate the solutions graphically.
In Fig.\ \ref{graph4}, we plot the functions $\dot{C}(\lambda w)$ and
$(\zeta-\rho/\lambda)\cdot w$ vs.\ $w$ for the following parameter values:
$\delta=0.95$, $q=5$, $z_0=0.5$, $\lambda=1$, and $\zeta-\rho/\lambda=0.13$.
As can be seen, in this example, there are two positive solutions (in addition to the trivial
solution, $w_0=0$), which are approximately, $w_1=3.71$ and $w_2=4.58$. Thus, in
this case, $\calS(\zeta)=\{0,3.71,4.58\}$, which corresponds to the set of
power levels, $\calS^2(\zeta)=\{0,13.7641,20.9764\}$. According to the above
discussion, optimal signaling is associated with time-sharing between two out
of these three signal levels. Given this simple fact, the optimal signal
levels, say, $a\ge 0$ and $b> 0$, and the optimal weight parameter, $\alpha$, can also be found directly by maximizing the MD
error exponent expression with respect to these parameters, 
subject to the power
constraint, $(1-\alpha)a^2+\alpha b^2=P_w$, similarly as was done earlier in the example
of the 4-ASK signal in Section \ref{optimalw}. 

\section{Detectors Based on Linear Combinations of Correlation and Energy}
\label{corr+energy}

In this section, we provide a brief outline of a possible extension of the scope to a broader
class of detectors that compare the test statistic
$$\sum_{t=1}^nw_tY_t+\alpha\sum_{t=1}^nY_t^2$$
to a threshold, $T=\theta n$. The motivation stems from the fact that 
the two hypotheses, $\calH_0$ and $\calH_1$, differ not only in
the presence of the signal, $\{s_t\}$, but also in the presence of the SIN, $\{Z_t\}$, which adds to the energy (or the variance) of
the received signal. In the extreme case, where $s_t\equiv 0$, the simple correlation detector we examined so far 
(corresponding to $\alpha=0$) would be useless, but still, 
one expects to be able to distinguish between the two 
hypotheses thanks to the different energies of the received signal. 
Indeed, if $\{Z_t\}$ is Gaussian white noise (Case 1), the optimal LRT detector obeys this structure with
$\alpha=\sigma_Z^2/[2(\sigma_N^2+\sigma_Z^2)]$. 

For practical reasons, it would also be relevant to consider detectors that are based on
$$\sum_{t=1}^nw_tY_t+\alpha\sum_{t=1}^n|Y_t|,$$
where the second term is another measure of the signal intensity, but with the advantage 
that its calculation does not require multiplications. We shall consider both classes of detectors, 
but provide merely the basic derivations of the MD exponent, without attempt to arrive at full, explicit solutions.
Nevertheless, we will make an attempt to make some observations on the solutions.

We begin with the first class of detectors mentioned above.
The FA probability is readily bounded by
\begin{eqnarray}
P_{\mbox{\tiny
FA}}(\theta)&=&\mbox{Pr}\left\{\sum_{t=1}^nw_tN_t+\alpha\sum_{t=1}^nN_t^2 \ge
\theta n\right\}\nonumber\\
&\le&\exp\left\{-n\sup_{\lambda\ge
	0}\left[\lambda\theta-\frac{\lambda^2\sigma_N^2P_w}{2(1-2\alpha\lambda\sigma_N^2)}+
	\frac{1}{2}\ln(1-2\alpha\lambda\sigma_N^2)\right]\right\},
\end{eqnarray}
which depends on $\bw$ only via $P_w$, as before.

As for the MD probability, we define
\begin{equation}
A=\frac{1}{n}\sum_{t=1}^n(w_ts_t+\alpha s_t^2)
\end{equation}
and
\begin{equation}
u_t=w_t+2\alpha s_t,~~~~~t=1,2,\ldots,n.
\end{equation}
Then,
\begin{eqnarray}
P_{\mbox{\tiny
MD}}(\theta)&=&\mbox{Pr}\left\{\sum_{t=1}^nw_t(s_t+Z_t+N_t)+\alpha\sum_{t=1}^n(s_t+Z_t+N_t)^2<\theta
n\right\}\nonumber\\
&=&\mbox{Pr}\left\{nA
+\sum_{t=1}^nu_t(Z_t+N_t)+\alpha\sum_{t=1}^n(Z_t+N_t)^2<n\theta\right\}\nonumber\\
&\le&\bE\left\{\exp\left[\lambda
n(\theta-A)-\lambda\sum_{t=1}^nu_t(Z_t+N_t)-\alpha\lambda\sum_{t=1}^n(Z_t+N_t)^2\right]\right\}\nonumber\\
&=&\bE\bigg\{\exp\left[\lambda
n(\theta-A)-\lambda\sum_{t=1}^nu_t(Z_t+N_t)\right]\times\nonumber\\
& &\prod_{t=1}^n\exp\left[-\alpha\lambda(Z_t+N_t)^2\right]\bigg\}\nonumber\\
&\eqa&\bE\bigg\{\exp\left[\lambda
n(\theta-A)-\lambda\sum_{t=1}^nu_t(Z_t+N_t)\right]\times\nonumber\\
& &\prod_{t=1}^n\left[(4\pi\alpha\lambda)^{-1/2}\int_{-\infty}^{\infty}\exp\left\{-jq_t(Z_t+N_t)-
\frac{q_t^2}{4\alpha\lambda}\right\}\mbox{d}q_t\right]\bigg\}\nonumber\\
&=&e^{\lambda
	n(\theta-A)}\prod_{t=1}^n\left[(4\pi\alpha\lambda)^{-1/2}\int_{-\infty}^{\infty}\bE\left\{\exp\left[-
(\lambda
u_t+jq_t)(Z_t+N_t)\right]\right\}\exp\left(-\frac{q_t^2}{4\alpha\lambda}\right)\mbox{d}q_t\right]\nonumber\\
&=&e^{\lambda
	n(\theta-A)}\prod_{t=1}^n\bigg[(4\pi\alpha\lambda)^{-1/2}\int_{-\infty}^{\infty}\bE\left\{\exp\left[-
(\lambda u_t+jq_t)Z_t\right]\right\}\times\nonumber\\
& &\bE\left\{\exp\left[-(\lambda
u_t+jq_t)N_t\right]\right\}\cdot\exp\left(-\frac{q_t^2}{4\alpha\lambda}\right)\mbox{d}q_t\bigg]\nonumber\\
&=&e^{\lambda
	n(\theta-A)}\prod_{t=1}^n\bigg[(4\pi\alpha\lambda)^{-1/2}\int_{-\infty}^{\infty}\bE\left\{\exp\left(-
\lambda u_tZ_t\right)e^{-jq_tZ_t}\right\}\times\nonumber\\
& &\exp\left(\frac{1}{2}\sigma_N^2
[\lambda u_t+jq_t]^2\right)\cdot\exp\left(-\frac{q_t^2}{4\alpha\lambda}\right)\mbox{d}q_t\bigg]\nonumber\\
&=&e^{\lambda
	n(\theta-A)}\prod_{t=1}^n\bigg[(4\pi\alpha\lambda)^{-1/2}\int_{-\infty}^{\infty}\bE\left\{\exp\left(-
\lambda u_tZ_t\right)e^{-jq_tZ_t}\right\}\times\nonumber\\
& &\exp\left(\frac{1}{2}\sigma_N^2\lambda^2u_t^2\right)e^{j\sigma_N^2\lambda
u_tq_t}
\cdot\exp\left\{-\left(\frac{\sigma_N^2}{2}+\frac{1}{4\alpha\lambda}\right)q_t^2\right\}\mbox{d}q_t\bigg]\nonumber\\
&=&\exp\bigg\{\lambda
	n(\theta-A)+\frac{1}{2}\sigma_N^2\lambda^2\sum_{t=1}^nu_t^2\bigg\}\times\nonumber\\
	& &\prod_{t=1}^n\bigg[(4\pi\alpha\lambda)^{-1/2}
\int_{-\infty}^{\infty}\bE\left\{\exp\left(-
\lambda u_tZ_t\right)\cos((Z_t-\sigma_N^2\lambda u_t)q_t)\right\}\times\nonumber\\
& &\exp\left\{-\left(\frac{\sigma_N^2}{2}+\frac{1}{4\alpha\lambda}\right)q_t^2\right\}\mbox{d}q_t\bigg],
\end{eqnarray}
where $j=\sqrt{-1}$ and (a) is due to the identity
\begin{equation}
e^{-a x^2}=
	(4\pi a)^{-1/2}\int_{-\infty}^{\infty}e^{-jqx}\exp\left\{-
\frac{q^2}{4a}\right\}\mbox{d}q,~~~~~a> 0,
\end{equation}
which is the characteristic function of a zero-mean Gaussian random variable with variance $2a$.\footnote{Alternatively, it
can be viewed as the Fourier transform relation between two Gaussians, one in the domain of $x$ and one in the domain of $q$.}
We now define
\begin{equation}
C_\alpha(v)\dfn\ln\left[\frac{1}{\sqrt{4\pi\alpha\lambda}}\int_{-\infty}^{\infty}\bE\left\{\exp\left(-
vZ\right)\cos((Z-\sigma_N^2v)q)\right\}\cdot
\exp\left\{-\left(\frac{\sigma_N^2}{2}+\frac{1}{4\alpha\lambda}\right)q^2\right\}\mbox{d}q\right],\nonumber
\end{equation}
and we arrive at the following expression of the MD exponent:
\begin{eqnarray}
	E_{\mbox{\tiny MD}}(\theta)&=&\sup_{\lambda\ge 0}
	\lim_{n\to\infty}\left\{\lambda(A-\theta)-\frac{1}{2}\lambda^2\sigma_N^2
\cdot\frac{1}{n}\sum_{t=1}^nu_t^2-\frac{1}{n}\sum_{t=1}^nC_\alpha(\lambda u_t)\right\}\nonumber\\
	&=&\sup_{\lambda\ge 0}\bigg\{\lambda\left(\bE\{S\cdot U\}-\alpha P_s-\theta\right)
-\frac{1}{2}\lambda^2\sigma_N^2
	\cdot\bE\{U^2\}-
	\bE\{C_\alpha(\lambda U)\bigg\},
\end{eqnarray}
where $U=W+2\alpha S$.
Note that this expression is of the same form of the one we had earlier, except that $W$ is replaced by
$U$, $\theta$ is replaced by $\theta+\alpha P_s$, and $C$ is
replaced by $C_\alpha$.\footnote{It is easy to verify that
$\frac{1}{2}\lambda^2\sigma_N^2u^2+C_\alpha(\lambda u)$ is convex in $u$ simply
because $\ln\bE\left\{
\exp\left[\lambda
(\theta+\alpha P_s)-su)-\lambda
u(Z+N)-\alpha\lambda(Z+N)^2\right]\right\}$ is such. Therefore, its
derivative is monotonically non-decreasing.}
This expression should now be jointly 
maximized w.r.t.\ $f_{US}$ subject to the power constraints,
$\bE\{S^2\}\le P_s$, $\bE\{(U-2\alpha S)^2\le P_w$. 
Using the same techniques as before, it is
not difficult to infer that the optimal $S$ for a given $U$ is linear in $U$, whereas the
optimal $U$ for a given $S$ is given by a non-linear equation. Whenever the number of simultaneous solutions to both
equations is finite, the signal levels can be optimized directly, as before. As for the optimization of $\alpha$,
among all pairs $\{(\alpha,P_w)\}$ 
that give rise to the same value of the FA exponent, one chooses the one that maximizes the MD exponent.

Moving on to the second class of detectors,
the analysis can be carried out using the
same technique as above, where the this time, we use the Fourier transform
identity,
\begin{equation}
	e^{-a|x|}=\frac{a}{\pi}\int_{-\infty}^{+\infty}\frac{e^{-jqx}\mbox{d}q}{q^2+a^2},~~~~a> 0
\end{equation}
which, as before, enables to exploit the independence between $Z_t$ and $N_t$ once the expectation operator is commuted
with the inverse Fourier transform integral over $q$.
Equipped with this identity, we have
\begin{eqnarray}
P_{\mbox{\tiny
MD}}(\theta)&=&\mbox{Pr}\left\{\sum_{t=1}^nw_t(s_t+Z_t+N_t)+\alpha\sum_{t=1}^n|s_t+Z_t+N_t|<\theta
n\right\}\nonumber\\
&\le&\bE\left\{\exp\left[\lambda\left(n\theta-\sum_{t=1}^nw_t(s_t+Z_t+N_t)-\alpha\sum_{t=1}^n|s_t+Z_t+N_t|\right)\right]\right\}\nonumber\\
&=&e^{\lambda
n\theta}\prod_{t=1}^n\left(\bE\left\{\exp\left[-\lambda w_t(s_t+Z_t+N_t)\right]\exp\left[-\alpha\lambda|s_t+Z_t+N_t|\right]\right\}\right)\nonumber\\
&=&e^{\lambda
n\theta}\prod_{t=1}^n\left(\bE\left\{\exp\left[-\lambda
w_t(s_t+Z_t+N_t)\right]\cdot\exp\left[-\alpha\lambda|s_t+Z_t+N_t|\right]\right\}\right)\nonumber\\
&=&e^{\lambda
n\theta}\prod_{t=1}^n\left(\bE\left\{\exp\left[-\lambda
w_t(s_t+Z_t+N_t)\right]\cdot\frac{\alpha\lambda}{\pi}\int_{-\infty}^{\infty}
\frac{e^{-jq_t(s_t+Z_t+N_t)}\mbox{d}q_t}{q_t^2+\alpha^2\lambda^2}\right\}\right)\nonumber\\
&=&\exp\left\{\lambda\left(
n\theta-\sum_{t=1}^nw_ts_t\right)\right\}\prod_{t=1}^n\left(\bE\left\{
\frac{\alpha\lambda}{\pi}\int_{-\infty}^{\infty}
\frac{e^{-jq_ts_t}e^{-(\lambda w_t+jq_t)(Z_t+N_t)}\mbox{d}q_t}{q_t^2+\alpha^2\lambda^2}\right\}\right)\nonumber\\
&=&\exp\left\{\lambda\left(
n\theta-\sum_{t=1}^nw_ts_t\right)\right\}\prod_{t=1}^n\left(
\frac{\alpha\lambda}{\pi}\int_{-\infty}^{\infty}
\frac{e^{-jq_ts_t}\bE\left\{e^{-(\lambda w_t+jq_t)(Z_t+N_t)}\right\}\mbox{d}q_t}{q_t^2+\alpha^2\lambda^2}\right)\nonumber\\
&=&\exp\left\{\lambda\left(
n\theta-\sum_{t=1}^nw_ts_t\right)\right\}\times\nonumber\\
& &\prod_{t=1}^n\left(
\frac{\alpha\lambda}{\pi}\int_{-\infty}^{\infty}
\frac{e^{-jq_ts_t}\bE\left\{e^{-(\lambda
w_t+jq_t)Z_t}\right\}\bE\left\{e^{-(\lambda w_t+jq_t)N_t}\right\}\mbox{d}q_t}{q_t^2+\alpha^2\lambda^2}\right)\nonumber\\
&=&\exp\left\{\lambda\left(
n\theta-\sum_{t=1}^nw_ts_t\right)\right\}\times\nonumber\\
& &\prod_{t=1}^n\left(
\frac{\alpha\lambda}{\pi}\int_{-\infty}^{\infty}
\frac{e^{-jq_ts_t}\bE\left\{e^{-(\lambda
w_t+jq_t)Z_t}\right\}\exp\left\{\frac{1}{2}(\lambda
w_t+jq_t)^2\sigma_N^2\right\}\mbox{d}q_t}{q_t^2+\alpha^2\lambda^2}\right)\nonumber\\
&=&\exp\left\{\lambda\left(
n\theta-\sum_{t=1}^nw_ts_t\right)+\frac{1}{2}\lambda^2\sigma_N^2\sum_{t=1}^nw_t^2\right\}\times\nonumber\\
& &\prod_{t=1}^n\left(
\frac{\alpha\lambda}{\pi}\int_{-\infty}^{\infty}
\frac{\bE\left\{e^{-\lambda
w_tZ_t}\exp\{jq_t(Z_t+\sigma_N^2\lambda w_t-s_t)\}\right\}
e^{-q_t^2\sigma_N^2/2}\mbox{d}q_t}{q_t^2+\alpha^2\lambda^2}\right)\nonumber\\
&=&\exp\left\{\lambda\left(
n\theta-\sum_{t=1}^nw_ts_t\right)+\frac{1}{2}\lambda^2\sigma_N^2\sum_{t=1}^nw_t^2\right\}\times\nonumber\\
& &\prod_{t=1}^n\left(
\frac{\alpha\lambda}{\pi}\int_{-\infty}^{\infty}
\frac{\bE\left\{e^{-\lambda
w_tZ_t}\cos((Z_t+\sigma_N^2\lambda w_t-s_t)q_t)\right\}
e^{-q_t^2\sigma_N^2/2}\mbox{d}q_t}{q_t^2+\alpha^2\lambda^2}\right).
\end{eqnarray}
Thus, defining
\begin{equation}
C_\alpha(v,s)=\ln\left[\frac{\alpha\lambda}{\pi}\int_{-\infty}^{\infty}
\frac{\bE\left\{e^{-v
Z}\cos((Z+\sigma_N^2v-s)q)\right\}
e^{-q^2\sigma_N^2/2}\mbox{d}q}{q^2+\alpha^2\lambda^2}
\right],
\end{equation}
the MD exponent is
\begin{equation}
E_{\mbox{\tiny MD}}(\theta)=\sup_{\lambda\ge
	0}\left\{\lambda\left(\bE\{W\cdot S\}-\theta\right)-\frac{1}{2}\lambda^2\sigma_N^2\bE\{W^2\}-
	\bE\{C_\alpha(\lambda W,S)\}\right\}.
\end{equation}
However, in this case, there is an additional complication, which stems from
the fact that the FA exponent depends on $\bw$ not only via $P_w$. 
A standard Chernoff-bound analysis yields
\begin{eqnarray}
E_{\mbox{\tiny FA}}(\theta)&=&\sup_{\lambda\ge
	0}\bigg(\lambda\theta-\frac{1}{2}\lambda^2\sigma_N^2(\bE\{W^2\}+\alpha^2)-\nonumber\\
	& &\bE\left\{\ln\left[e^{\lambda\alpha
W}\left[1-Q\left(\frac{\lambda(W+\alpha)}{\sigma}\right)\right]+e^{-\lambda\alpha
	W}Q\left(\frac{\lambda(W-\alpha)}{\sigma}\right)\right]\right\}\bigg).
\end{eqnarray}
Therefore, the maximization of the MD exponent will have to incorporate the full asymptotic PDF of $W$ and not just its second moment.


\end{document}